\documentclass{llncs}
\usepackage{epsfig}

\usepackage{amsmath}
\usepackage{amsfonts}
\usepackage{amssymb}
\newtheorem{assumption}[theorem]{Assumption}
\begin{document}

\pagestyle{headings}

\title{Numerical Simulations of Electrohydrodynamics flow model based on Burgers' Equation with Transport of Bubbles}
\author{J\"urgen Geiser and Paul Mertin}
\institute{Ruhr University of Bochum, \\
The Institute of Theoretical Electrical Engineering, \\
Universit\"atsstrasse 150, D-44801 Bochum, Germany \\
\email{juergen.geiser@ruhr-uni-bochum.de}}
\maketitle

\begin{abstract}

In this paper we present numerical models for electrodynamical flows with time-dependent electrical
fields with transport of bubbles. Such models are applied in e-jet printing, e.g., additive manufacturing (AM), and
convective cooling, electrostatic precipitation, plasma assisted combustion.

We present a coupled model, consisting of two models. The first model is the underlying hydrodynamical simulation model,
which is based on the Burgers' equation. The second model is a transport model based on level-set equations, which 
transport the bubbles in the flow field. 
We derive the modeling of the two coupled modeling equations and discuss decoupled and coupled versions
for solving such delicate problems.
In the numerical analysis, we discuss the solvers methods, which are based on splitting approaches and level-set
techniques. In the first numerical results, we present decoupled and coupled model simulations of the
bubbles in the liquid flow.

\end{abstract}



\section{Introduction}

We are motivated to simulate bubble transport and their formation in liquid,
based on E-fields and different flow-fields.
Such formation and transport of gas-bubbles in liquid are important 
for the controlled production in  additive manufacturing (AM), convective cooling, electrostatic precipitation, plasma assisted combustion,
see \cite{gu2011}, \cite{kumar1970}, \cite{yang2007} and \cite{hayashi2015}.

For simulating such processes, we are motivated to extend an underlying hydrodynamics model,
which is based on the Burgers' equation with an electrodynamical equation and 
additional a two-phase equation, which models the bubble transport and modifications, see \cite{behe2019}.

We consider the single models and coupled them into a multi-physics model based on the different modeled
processes, i.e., flow-field model, E-field model and phase field model.
Here, the numerical solvers are important, while we deal with different
scale-dependent models, see \cite{geiser2011} and \cite{geiser_2016}.

We present splitting methods, which separate each modeling equation and solve them with the
optimal solver for each equations, see \cite{geiser_2018_2}.

We apply a modified model of the surface tension, see \cite{sussmann2009}, which is important for the formation and transport of the bubbles
in liquids, see \cite{simmons2015}
and \cite{geiser2018}. We present first numerical results with decoupled and coupled flow-fields.

The paper is organized as following:
The modeling problems and their solvers are presented in Section \ref{modell}.
The solver of the models are discussed in Section \ref{solver}
The numerical experiments are presented in Section \ref{exp}.
In the contents, that are given in Section \ref{concl}, 
we summarize our results.

\section{Mathematical Model}
\label{modell}

The modeling is based on an electrohydrodynamics model coupled with phase field models.

We have the following coupled modeling equations:
\begin{itemize}
\item Electrical field equation
\item Phase-field equation
\item Fluid-flow equation
\end{itemize}

\subsection{Modeling: Electrical field equations}

We deal with the electrohydordynamics. We have given the Poisson equation as:
\begin{eqnarray}
\nabla \cdot (\epsilon_0 \; \epsilon_r \nabla \phi) = - \rho_e ,
\end{eqnarray}
$\epsilon_0$ is the permittivity of vacuum, $\epsilon_r$ the dielectric constant,
$\phi$ the potential and $\rho_e$ the free charge density.

The free charge conservation is given as:
\begin{eqnarray}
\frac{\partial \rho_e}{\partial t} + {\bf u} \cdot \nabla \rho_e = - \nabla \cdot ( \sigma \nabla \rho_e)  ,
\end{eqnarray}
where $\sigma$ is the electrical conductivity, ${\bf u}$ the velocity field and ${\bf E} = - \nabla \phi$ is the electrical field.
While the second term on the left hand side presents the convection of the free charges
and the right-hand side term presents the transport by electromigration.

\subsection{Modeling: Phase field equations}

We deal with the volume-fraction $c$, which satisfies the advection equation:
\begin{eqnarray}
\frac{\partial c}{\partial t} + {\bf u} \cdot \nabla c = 0  ,
\end{eqnarray}
where ${\bf u}$ is the velocity field and $c=1$ is the bubble/drop phase, while $c=0$
is the bulk fluid phase.

\begin{remark}
The full phase field equation is given by the Cahn-Hillard equation, while the parameter
$c$ is the phase parameter.
\end{remark}

\subsection{Modeling: Fluid-flow equation}

We deal with the fluid-flow equation, which satisfy the Navier-Stokes equation:
\begin{eqnarray}
&& \rho \frac{\partial {\bf u}}{\partial t} + \rho ( {\bf u} \nabla ) {\bf u} = - \nabla p + \mu \nabla^2 {\bf u} + \rho {\bf g} + {\bf f}_{sf} + {\bf f}_{ef}  , \\
&& \nabla \cdot {\bf u} = 0 ,
\end{eqnarray}
where ${\bf u}$ is the velocity field and $p$ is the pressure, $\mu$ is the dynamic viscosity, $\rho$ is the fluid density, ${\bf f}_{sf}$ is the surface tension, ${\bf f}_{ef}$ 
is the electrostatic force, ${\bf g}$ is the gravitational acceleration.

The forces are given as:
\begin{eqnarray}
&& {\bf f}_{sf} = \sigma \kappa(\nabla c) , \\
&& {\bf f}_{ef} = \nabla \tau_e = \rho_e {\bf E} - \frac{1}{2} |{\bf E}|^2 \nabla \epsilon,
\end{eqnarray}
where $\sigma$ is the surface tension, $\kappa$ is the interface curvature with $\kappa(\nabla c) = \nabla \cdot \frac{\nabla c}{|\nabla c|}$ and $\tau_e$ is the Maxwell stress tensor.

\section{Solvers and splitting approaches}
\label{solver}

\begin{figure}[ht]
\begin{center}  
\includegraphics[width=6.0cm,angle=-0]{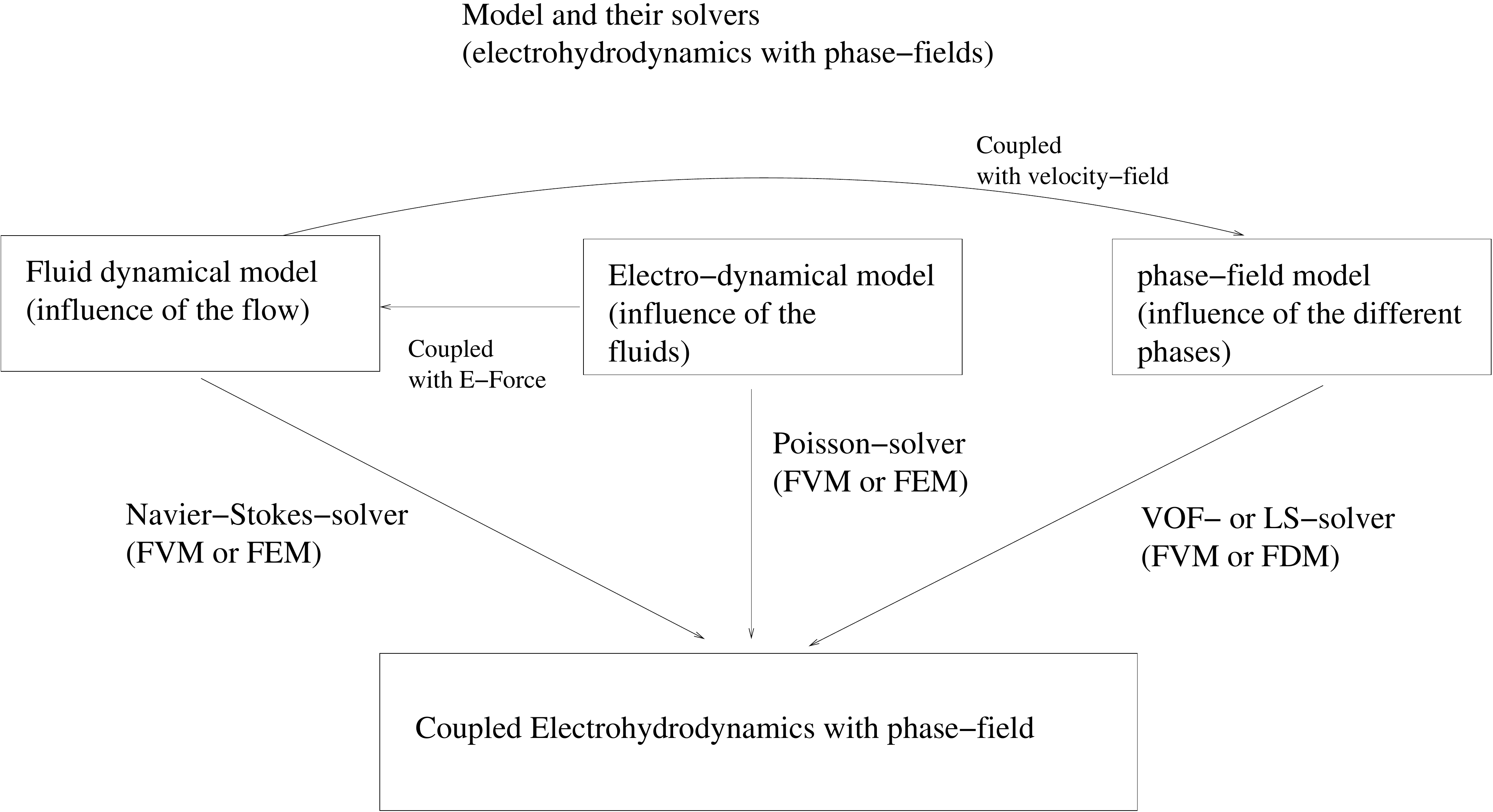}
\end{center}
\caption{\label{cycle} Numerical cycle of the near-far-field coupling.}
\end{figure}

We deal with the modeling and solver-ideas, as presented in Fig. \ref{cycle}.

\subsection{Reinitialization of the Level-Set method}

Based on the problem of the conservation of the volume of the
level-set method, we deal with the improvement of the
reinitialisation, see \cite{Radams2019}.

We have the following algorithm of the reinitialization:
\begin{enumerate}
  \item We initialize with $t=0$ and $c(0) = c_{init}$. we apply $n=1$ and we have the indicator function of the density $\rho(t_0) = \rho_{init} = 1.0$ and the volume of the bubble with  $V(t_0) = V_{init}$ .
  \item We compute with the next level-set step with time-step $\Delta t = t_n - t_{n-1}$. We have the old concentration at time $t_{n-1}$ with $c(t_{n-1})$ and we obtain the new concentration $c(t_n)$ at $t_n$.
  \item We compute the volume of the bubble $V(t_n)$ in the new time-point $t_n$.
  \item We apply the correction, which is based on the factor:
    \begin{eqnarray}
      \rho(t_n) = \rho(t_{n-1}) \frac{V(t_{n-1})}{V(t_n)} ,
      \end{eqnarray}

\item We apply the reinitialization of the concentration, which is given as:
    \begin{eqnarray}
      c_{reinit}(t_n) = \rho(t_n) c(t_n).
      \end{eqnarray}

  \item We set  $c(t_n) = c_{reinit}(t_n)$ and apply $n = n+1$, if $n+1 = N+1$ we are done, else we go to Step $2$.

    \end{enumerate}

\begin{figure}[ht]
\begin{center} 
\includegraphics[width=12.0cm,angle=-0]{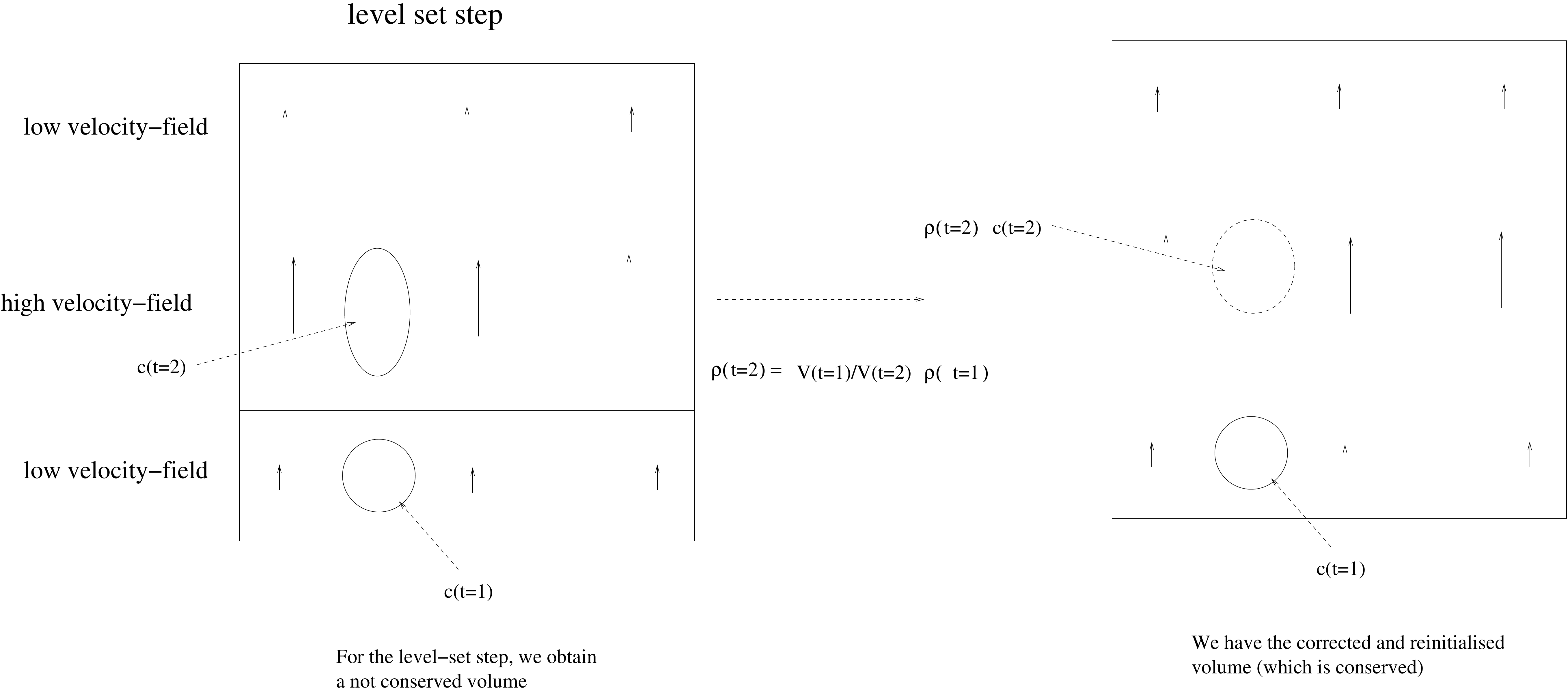}
\end{center}
\caption{\label{reinit} Reinitialization in the level set step.}
\end{figure}

The idea of the algorithm is given in the Figure \ref{reinit}.

\subsection{Solver methods for the Burgers' and Navier-Stokes equation}
\label{burger-ns}

In the following, we discuss the discretization of the Burger's and NS equation.

\subsubsection{Discretization of the Burgers' equation}

We apply the following splitting approach to the 2D Burgers' equation, which is given as:
\begin{eqnarray}
&& \partial_t u = - u u_x - v \; u_y +  D \; (u_{xx} + u_{yy}) + f_{el, x}(t),  \\
&& \partial_t v = - u \; v_x - v  \; v_y +  D \; (v_{xx} + v_{yy}) + f_{el, y}(t),  
\end{eqnarray}
with initial and boundary conditions.

We rewrite into the operator-notation as:
\begin{eqnarray}
&& \partial_t U = A(U) + B U + F_{el} ,
\end{eqnarray}
where $A$ is the nonlinear term (flow-term), $B$ is the linear viscosity-term
and $F_{el}$ is the right-hand side electrical field term.

We apply the following splitting approach:

\begin{itemize}
\item A-Step (Burgers' Step) with explicit time-discretization:

We apply the finite difference or finite volume scheme for the viscous Burgers' equation
and deal with the following discretized differential equations:

We apply the following splitting method with the time-step size $\Delta t = t^{n+1} -t^n$:
\begin{eqnarray}
&& U_1^{n+1} = U^n_1 - (U_1^n \nabla) U_1^n , \; U_1(t^n) = U(t^{n}) , \\ 
\end{eqnarray}
where we discretize the spatial operator with upwinding methods.

\item B-Step (Diffusion Step) with implicit time-discretization:

We apply the finite difference or finite volume scheme for the Diffusion equation
and deal with the following  discretized differential equations:

We apply the following splitting method with the time-step size $\Delta t = t^{n+1} -t^n$:
\begin{eqnarray}
  && U_2^{n+1} = U_2^n + B U_2^{n+1} , \; U_2(t^n) = U_1(t^{n+1}) , \\ 
  && U_2^{n+1} = (I - B)^{-1} U_2^n , \; U_2(t^n) = U_1(t^{n+1}) , 
\end{eqnarray}
where $B$ is the semi-discretized operator of the diffusion operator.

\item C-Step (RHS-Step)

We apply the RHS-operator
\begin{eqnarray} 
&&  U_3^{n+1} = U_3^{n} + \Delta t \; F^{n+1} , \; U_3(t^n) =  U_2(t^{n+1}) ,
\end{eqnarray}
where we apply the rhs with $F^{n+1} = (f_{el,x}(t^{n+1}), f_{el,y}(t^{n+1}))^t$.
We obtain the solution of the Burgers' equation $U(t^{n+1}) =  U_3^{n+1}$.
Then, we could go on to the first step.
\end{itemize}

\subsubsection{Discretization of the NS equation}
\label{ns_splitting}

We apply the following splitting approach to the Navier-Stokes equation.

We deal with the 2D NS equation, which is given as:
\begin{eqnarray}
&& \partial_t u + p_x = - (u^2)_x - (u v)_y +  \frac{1}{Re} (u_{xx} + u_{yy}),  \\
&& \partial_t v + p_y = - (u v)_x - (v^2)_y +  \frac{1}{Re} (v_{xx} + v_{yy}), \\
&& u_x + v_y = 0 ,  
\end{eqnarray}
with initial and boundary conditions.

We rewrite into the operator-notation as:
\begin{eqnarray}
&& \partial_t U = A(U) + B U + C P , \\
&& \nabla \cdot U = 0 ,  
\end{eqnarray}
where $A$ is the nonlinear term (flow-term), $B$ is the linear viscosity-term
and $P$ is the pressure-term.

We apply the following splitting approach:

\begin{itemize}
\item A-Step (Burgers' Step): \\
We apply the finite difference or finite volume scheme for the viscous Burgers' equation
and deal with the following  semi-discretized differential equations:

We apply the following splitting method with the time-step size $\Delta t = t^{n+1} -t^n$:
\begin{eqnarray}
&& \partial_t U_1 = A(U_1) + B U_1 , \; U_1(t^n) = U(t^{n}) , \\ 
\end{eqnarray}
where $A$ is the semi-discretized operator of the nonlinear convection operator of the NS equation
and $B$ is the semi-discretized operator of the diffusion operator of the NS equation.

\item B-Step (Correction-Step)

We apply the pressure term in the NS equation, which is solved as following: \\

We apply an implicit formulation as:
\begin{eqnarray} 
&& \frac{1}{\Delta t} (U_2^{n+1} - U_2^n) = - \nabla P^{n+1} , \; U_2(t^n) =  U_1(t^{n+1}) ,
\end{eqnarray}
where we apply the divergence of the solution and we have $ \nabla \cdot U_2^{n+1} = 0$.
Then, we reformulate with an additional derivation into the following Poisson equation:
\begin{eqnarray} 
&& - \Delta P^{n+1} = - \frac{1}{\Delta t} (\nabla \cdot U_2^n) .
\end{eqnarray}
Therefore we solve:
\begin{eqnarray} 
&& F^n = \nabla \cdot U_2^n , \; \mbox{Computation of the right hand side}, \\
&& - \Delta P^{n+1} = F^n , \; \mbox{Solving the Poisson's equation}, \\
&& G^{n+1} = \nabla P^{n+1} , \; \mbox{Computation of the pressure-gradient}, \\
&& U_2^{n+1} = U_2^n - \Delta t \; G^{n+1}, \; \mbox{Correction of the velocity field},
\end{eqnarray}
where we obtain the solution of the NS equation $U(t^{n+1}) =  U_2^{n+1}$.
Then, we could go on to the first step.
\end{itemize}

\section{Numerical Experiments}
\label{exp}

In the next subsections, we deal with the different
test-experiments based on the coupled equations.

We apply the following test- and Simulation-Software, that 
allows a flexibilisation of the software with respect to the EHD-model:
\begin{figure}[ht]
\begin{center} 
\includegraphics[width=8.0cm,angle=-0]{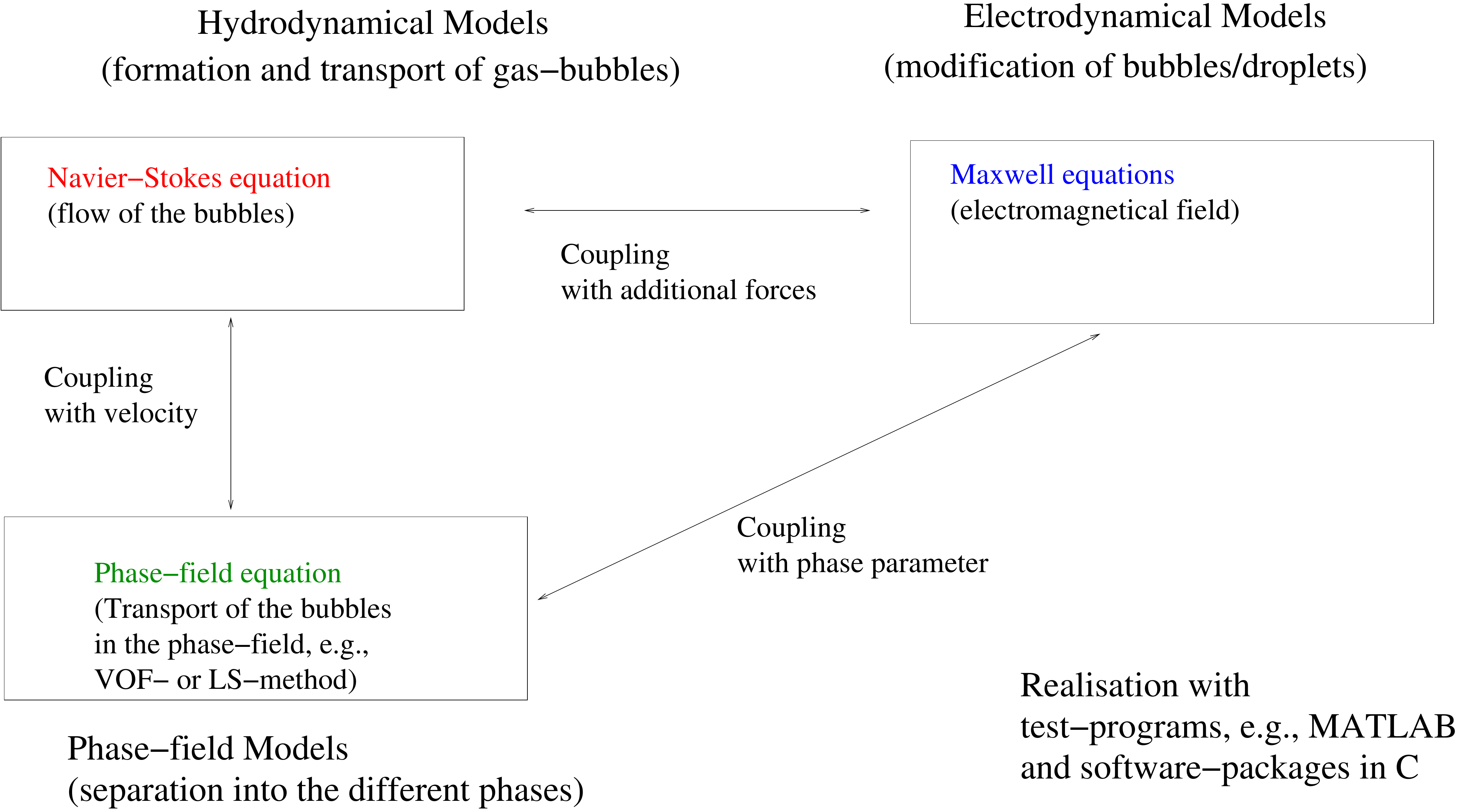}
\end{center}
\caption{\label{part_1_b} Software-flexibilisation.}
\end{figure}

\subsection{Test example 1: Slow and fast velocity fields}

In the following, we deal with the modeling-equation:

The modeling equations of the gas-bubble in the liquid is given by a transport equation,
which is based on the idea of the volume of fluid and level set methods.

We assume the indicator functions $u_i({\bf x}, t)$ with $i = 1, \ldots, M$, while $M$ is the number of bubble-sources.

We define the values of the function as:
\begin{eqnarray}
\left\{ \begin{array}{cc}
u_i({\bf x}, t) = 1 , & \mbox{gas-phase} , \\ 
u_i({\bf x}, t) = 0 , &  \mbox{liquid-phase} , \\ 
0 < u_i({\bf x}, t) < 1 , &  \mbox{interface} . 
\end{array} \right.
\end{eqnarray}

Furthermore, we assume the absence of phase changes and then, we deal with the
transport equation of the gas-phase as:
\begin{eqnarray}
\label{gas_phase_1_1}
&& \frac{\partial u_i}{\partial t} + \nabla({\bf v} \; u_i) = 0 ,  \;  (x,y, t) \in \Omega \times [0, T] , \; i = 1,  \ldots , M , \\
\label{gas_phase_2_1}
&& u_i({\bf x}, 0) = u_{i,0}({\bf x}) , \\
\label{gas_phase_3_1}
&& i = 1,  \ldots , M ,
\end{eqnarray}
where ${\bf v}$ is the velocity of the two-phase flow equation, and $c_i$ are the indicator functions of the gas-phases
and $u_{i,0}({\bf x})$ are the initial conditions.

\begin{assumption}
\label{ass_1}

We deal with the following assumption to reduce the computational amount of the far-field model:
\begin{itemize}
\item We can assume a constant velocity in the reactor. 
\item We assume that the water and the gas are not interacting or change the constant velocity.
\item We have such small velocities, such that the transport equations of the gas-bubbles are sufficient, see \cite{tsui2017}. 
\item We deal with a space and time-dependent velocity: ${\bf v} = {\bf v}({\bf x}, t)$ in the transport equations (\ref{gas_phase_1_1})-(\ref{gas_phase_3_1}),
  further, we also fulfill the CFL-condition of the explicit scheme.  with $\Delta t \le \frac{\Delta x}{\max_{({\bf x}, t) \in \Omega \times [0, T]} \{ |{\bf v}({\bf x}, t)| \}}$.
\item The concentration of the bubbles are given as $c_i({\bf x}, t) = c_{i,0}({\bf x}) u_i({\bf x}, t)$, while $c_{i,0}({\bf x})$ is the initial concentration of the 
bubble.
\end{itemize}

\end{assumption}

We study 2 different examples:

\begin{enumerate}

  \item Example 1: Two layer problem:

    We have the following velocity field for the bubble-transport, see Figure \ref{fig_bubble}.
    For the low velocity-field we have $v = (1,0)^t$ for the high velocity-field, we have $v=(2,0)^t$, as shown in Fig. \ref{fig_bubble}.
\begin{figure}[ht]
\begin{center} 
\includegraphics[width=6.0cm,angle=-0]{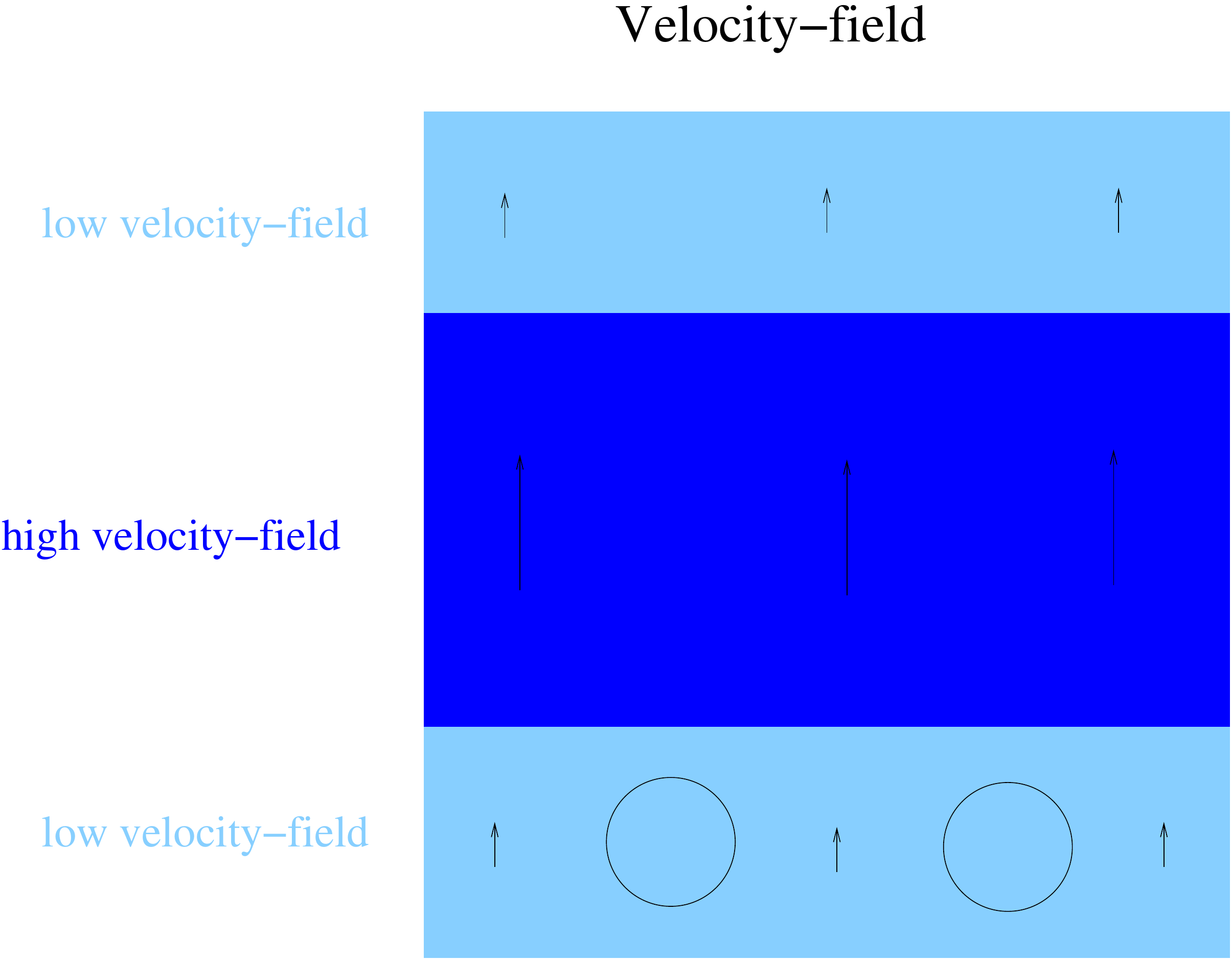}
\end{center}
\caption{\label{fig_bubble} Velocity-field to produce thin and large bubbles.}
\end{figure}

The results are computed with a MATLAB-code and are given in see Figure \ref{fig_bubble_2}.
\begin{figure}[ht]
\begin{center} 
\includegraphics[width=3.0cm,angle=-0]{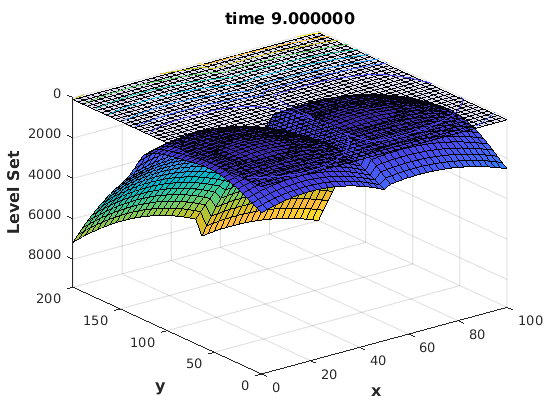}
\includegraphics[width=3.0cm,angle=-0]{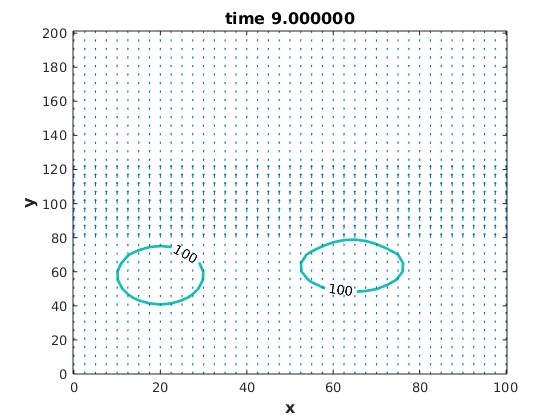} \\
\includegraphics[width=3.0cm,angle=-0]{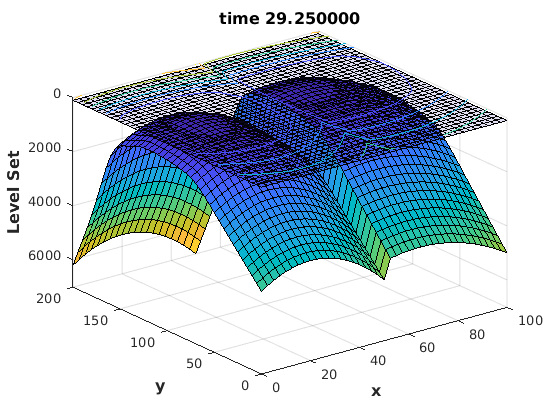}
\includegraphics[width=3.0cm,angle=-0]{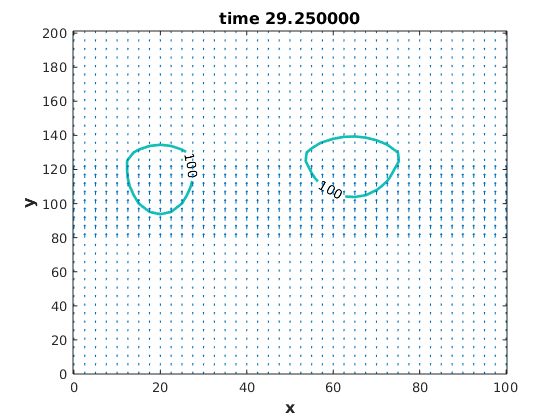} \\
\includegraphics[width=3.0cm,angle=-0]{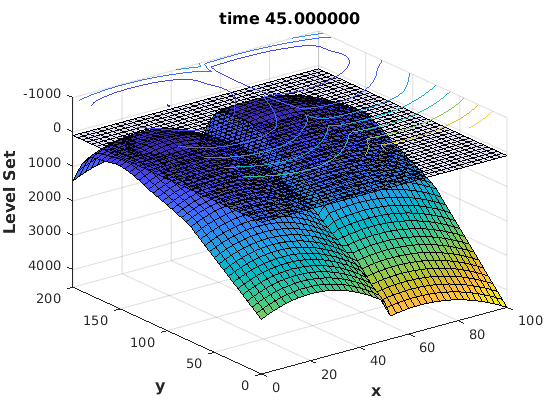}
\includegraphics[width=3.0cm,angle=-0]{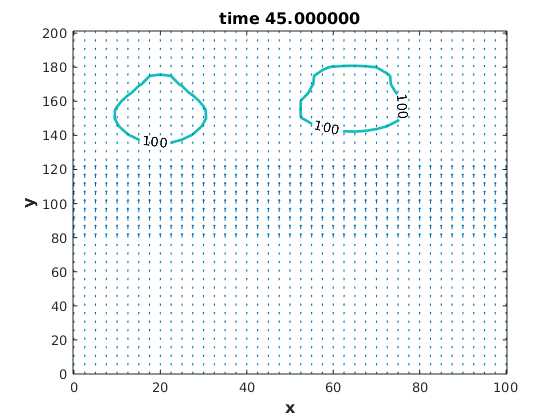}
\end{center}
\caption{\label{fig_bubble_2} Transport of 2 bubbles in a velocity field with different layers (low velocity-field with $v = 1.0$ and high velocity field with $v=2$), the level-set function (left figures) and the contour functions (right figures) are given at at time $t = 10$, $t=30$ and $t=45$ (from top to bottom).}
\end{figure}

 \item Example 2: Multi-layer problem:

In the following, we deal with five different layers, while we could also extend the number of layers. 
We have the following velocity field for the bubble-transport, see Figure \ref{fig_bubble_2_1}.
For the lowest velocity-field we have $v = (1,0)^t$, for the low velocity-field, we have $v=(2,0)^t$ and for the high velocity-field, we have $v=(3,0)^t$.
\begin{figure}[ht]
\begin{center} 
\includegraphics[width=5.0cm,angle=-0]{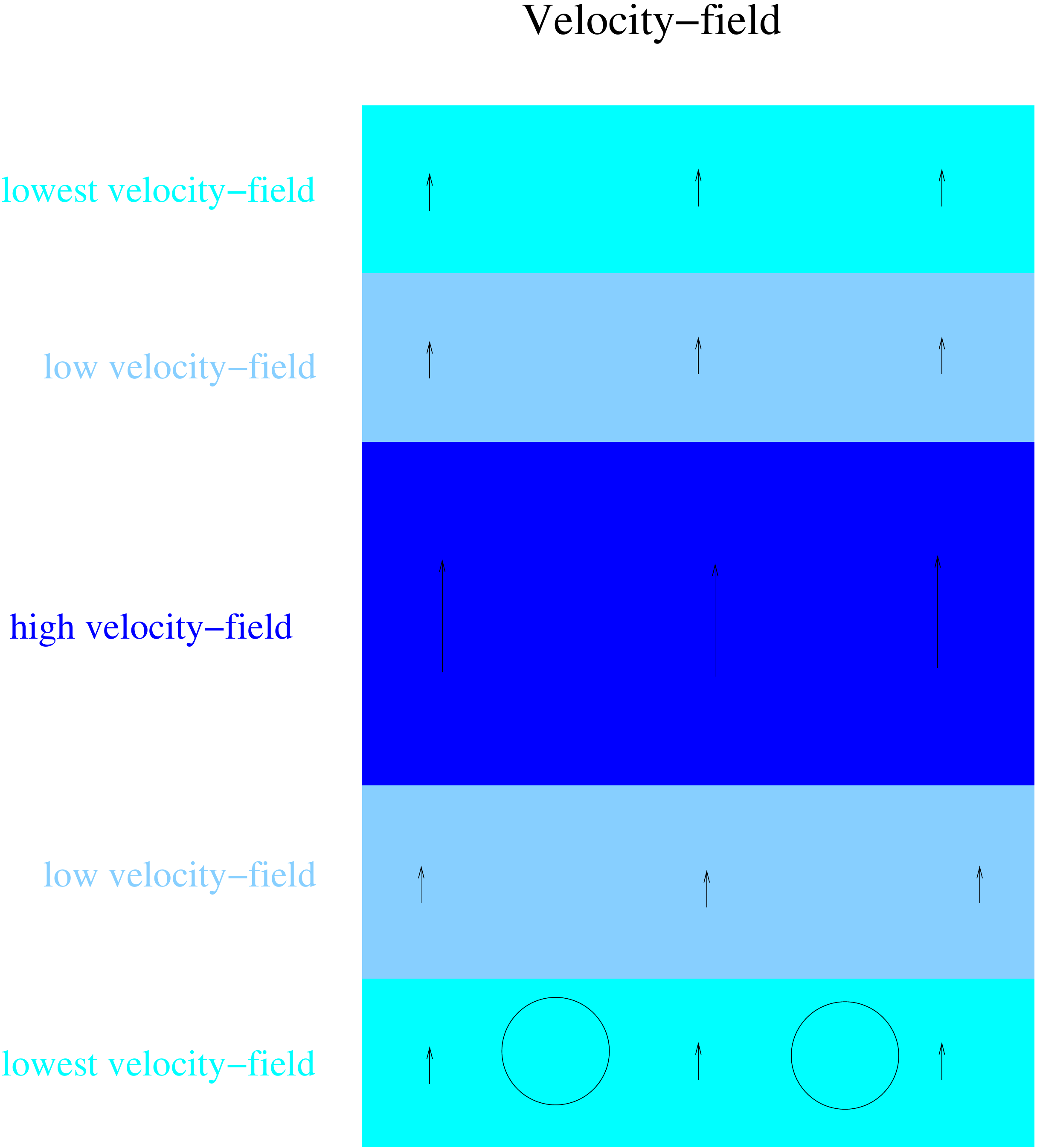}
\end{center}
\caption{\label{fig_bubble_2_1} Velocity-field to produce thin and large bubbles in a multi-layer field with different velocities.}
\end{figure}

The results are computed with the MATLAB-code and are given in see Figure \ref{fig_bubble_3}.
\begin{figure}[ht]
\begin{center} 
\includegraphics[width=3.0cm,angle=-0]{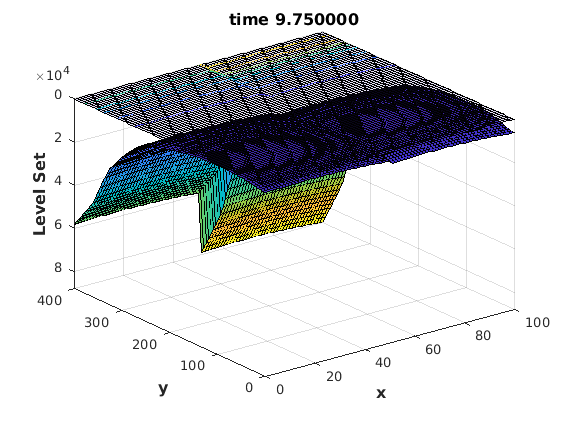}
\includegraphics[width=3.0cm,angle=-0]{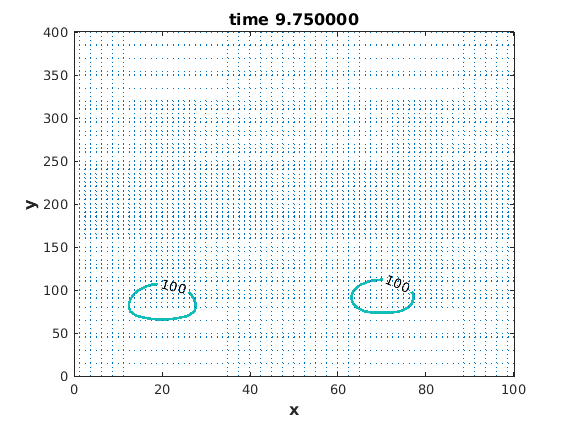} \\
\includegraphics[width=3.0cm,angle=-0]{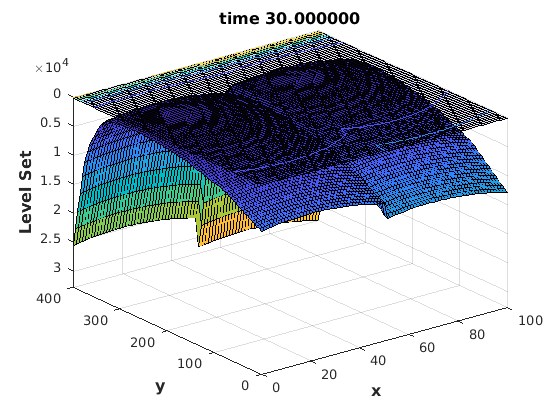}
\includegraphics[width=3.0cm,angle=-0]{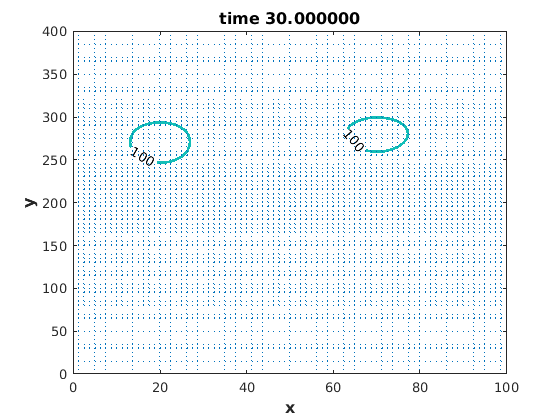} \\
\includegraphics[width=3.0cm,angle=-0]{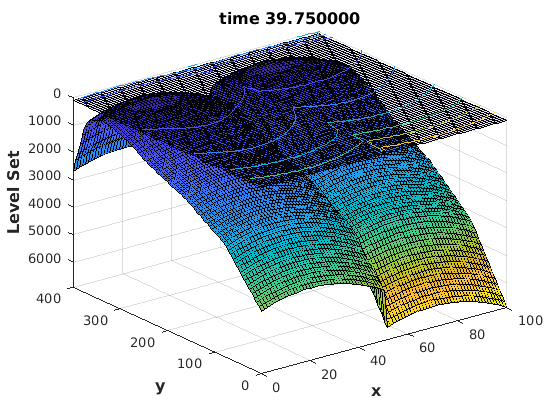}
\includegraphics[width=3.0cm,angle=-0]{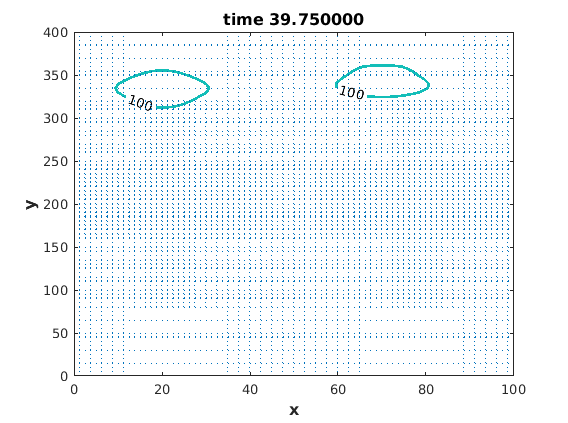}
\end{center}
\caption{\label{fig_bubble_3} Transport of 2 bubbles in a velocity field with 5 different layers (low $v = 1.0$, medium $v=2$ and high velocity field with $v=3$), the level-set function (left figures) and the contour functions (right figures) are given at at time $t = 10$, $t=30$ and $t=40$ (from top to bottom).}
\end{figure}

\end{enumerate}

\subsection{Test example 2: Decoupled EHD-model with Burgers' equation}

We apply the following decoupled model, which is applied with $p = 0$:
\begin{eqnarray}
&& \rho \frac{\partial {\bf u}}{\partial t} + \rho ( {\bf u} \nabla ) {\bf u} = \mu \nabla^2 {\bf u} + {\bf f}_{ef}  , {\bf x} \in [0, 100] \times [0, 400], t \in [0, T_{end}] , \\
&& {\bf u}({\bf x}, 0) = \left\{
\begin{array}{cc}
1.5 , &  12.5 \le x \le 56.25 \; \mbox{and} \;   150.0 \le x \le 225.0 , \\
1.0 , & \mbox{else}
\end{array}
\right.
\end{eqnarray}
where ${\bf u}$ is the 2D-velocity field and $p$ is the pressure, $\mu$ is the dynamic viscosity, $\rho$ is the fluid density, ${\bf f}_{ef}$ is the electrostatic force.
We apply $\rho = 1.0$ and $\mu = 0.01$ and $T_{end} = 40$.

The forces are given as:
\begin{eqnarray}
&& {\bf f}_{ef} = \nabla \tau_e = \rho_e {\bf E} ,
\end{eqnarray}
where we have ${\bf E} ={\bf E}_1 + {\bf E}_2$ , \\
$ {\bf E}_1(x,y) = \left\{ \begin{array}{cc}
  (\sin(4 \pi t/T), \sin(4 \pi t/T))^t, &  \mbox{for} \; (x,y) \in (0.45, 0.55) \times (0.45, 0.55) \\
  0 & \mbox{else}
\end{array} \right. $ and \\

$ {\bf E}_2(x,y) = \left\{ \begin{array}{cc}
  -(\sin(4 \pi t/T), \sin(4 \pi t/T))^t, &  \mbox{for} \; (x,y) \in (0.75, 0.85) \times (0.75, 0.85) \\
  0 & \mbox{else}
\end{array} \right. $ .

The velocity field is applied in the phase field model, which is given as:
\begin{eqnarray}
\label{gas_phase_1}
&& \frac{\partial \tilde{c}_i}{\partial t} + \nabla({\bf u} \; \tilde{c}_i) = 0 ,  \;  (x,y, t) \in \Omega \times [0, T] , \; i = 1,  \ldots , M , \\
\label{gas_phase_2}
&& \tilde{c}_i({\bf x}, 0) = \tilde{c}_{i,0}({\bf x}) , \\
\label{gas_phase_3}
&& i = 1,  \ldots , M ,
\end{eqnarray}
where ${\bf u}$ is the velocity of the two-phase flow equation, and $\tilde{c}_i$ are the indicator functions of the gas-phases
and $\tilde{c}_{i,0}({\bf x})$ are the initial conditions. The indicator functions $\tilde{c}_i({\bf x}, t)$ is given as:
\begin{eqnarray}
\left\{ \begin{array}{cc}
\tilde{c}_i({\bf x}, t) = 1 , & \mbox{gas-phase} , \\ 
\tilde{c}_i({\bf x}, t) = 0 , &  \mbox{liquid-phase} , \\ 
0 < \tilde{c}_i({\bf x}, t) < 1 , &  \mbox{interface} ,
\end{array} \right.
\end{eqnarray}
with $i = 1, \ldots, M$, while $M$ is the number of bubble-sources.

Further, we have the CFL-condition of the Burgers' Equation and the 
level-set equation for the
time-interval $[t^n, t^{n+1}]$ is given as:

\begin{itemize}
\item CFL-Condition for the Burgers' equation:
\begin{eqnarray}
&& \Delta t \le \frac{1}{\frac{\max\{{\bf u}^n\}}{\Delta x} + \frac{2 D }{\Delta x}} ,
\end{eqnarray}
where $\Delta t$ is the time-step, $\Delta x = \Delta y$ is the spatial step and
$\max\{{\bf u}^n\}$ is the absolute value of the velocity-field in time-step $t^n$ and $D$ is the diffusion parameter.

\item CFL-Condition for the Level-set equation:
\begin{eqnarray}
&& \Delta t \le \frac{\Delta x}{\max\{{\bf u}^n\}} ,
\end{eqnarray}
where $\Delta t$ is the time-step, $\Delta x = \Delta y$ is the spatial step and
$\max\{{\bf u}^n\}$ is the absolute value of the velocity-field in time-step $t^n$.

\end{itemize}

\begin{itemize}

\item EHD-model without E-field, we have $\rho_e = 0.0$: \\

The velocity field is computed by the Burgers' equation in advance at the beginning, see the Figures \ref{burger_no-E_de} (here without the E-field). Further, we apply the computed velocity field based on the decoupled Burgers' equation (without the E-field) into the phase field model,
which is computed with modified level-set method, see the Figures \ref{burger_no-E_de}.
\begin{figure}[ht]
\begin{center} 
\includegraphics[width=4.0cm,angle=-0]{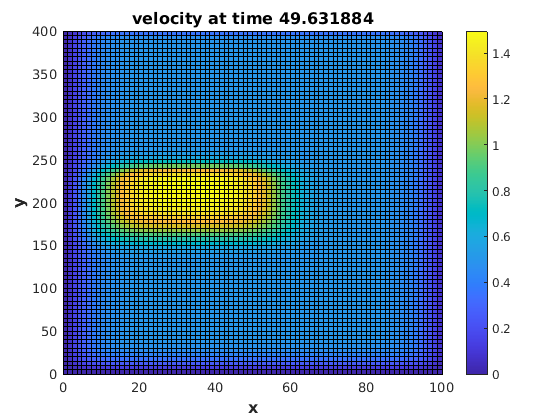}
\includegraphics[width=4.0cm,angle=-0]{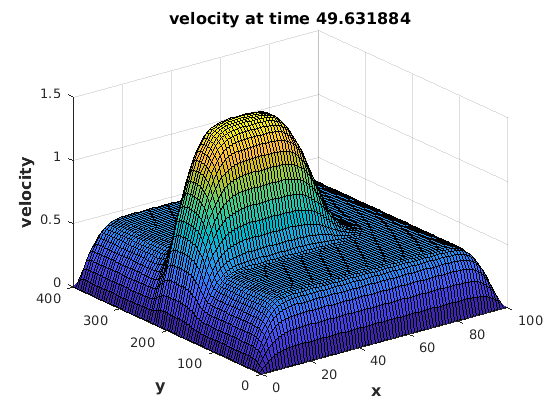} \\
\includegraphics[width=4.0cm,angle=-0]{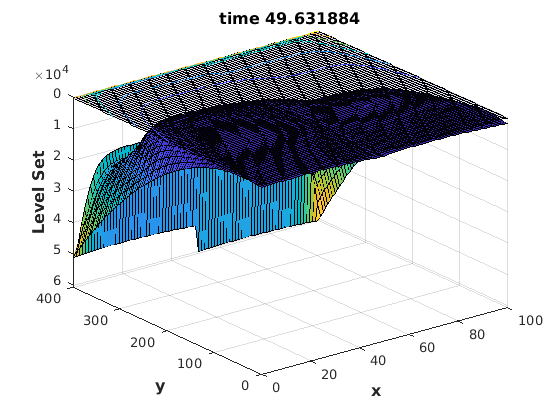}
\includegraphics[width=4.0cm,angle=-0]{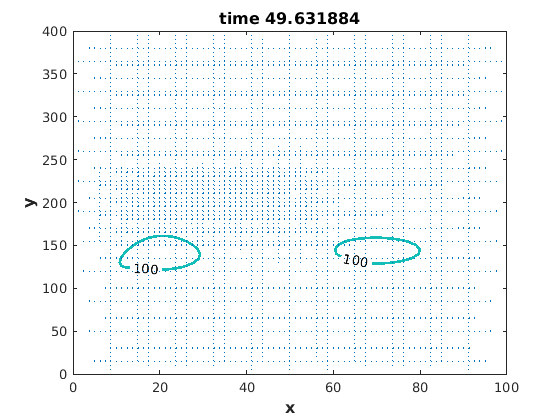} 
\end{center}
\caption{\label{burger_no-E_de} Upper figures: Velocity-field of the Burgers' equation without E-field in contour visualization (upper left figure) and velocity-field of the Burgers' equation without E-field in vectorial-visualization (upper right figure). Lower Figures: Transport of 2 bubbles in a velocity field with decoupled velocity field computed by the Burgers' equation without an E-field, the level-set function is given in the lower left figure and the contour functions is given in the lower right figure at time $t=50$.}
\end{figure}

\item EHD-model with E-field:, we have $\rho_e = 0.01$ \\

The velocity field is computed by the Burgers' equation and is computed in previous at the beginning, while we also apply the maximum amplitude of the E-field, see the Figures \ref{burger_E_de} (here with the E-field). 
Further, we apply the computed velocity field based on the decoupled Burgers' equation (with the E-field) into the phase field model,
which is computed with modified level-set method, see the Figures \ref{burger_E_de}.
\begin{figure}[ht]
\begin{center} 
\includegraphics[width=4.0cm,angle=-0]{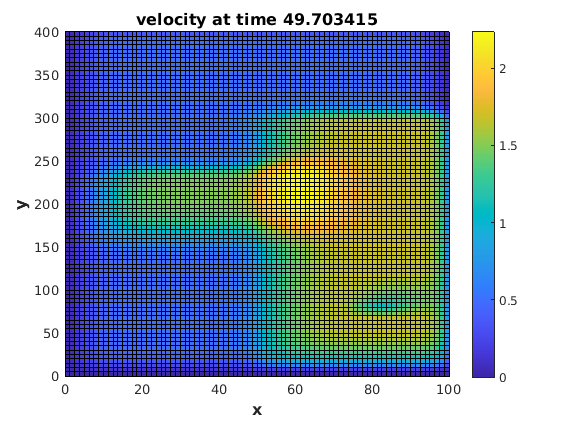}
\includegraphics[width=4.0cm,angle=-0]{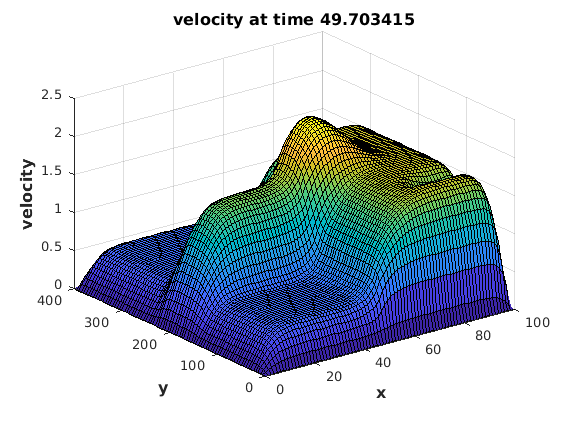} \\
\includegraphics[width=4.0cm,angle=-0]{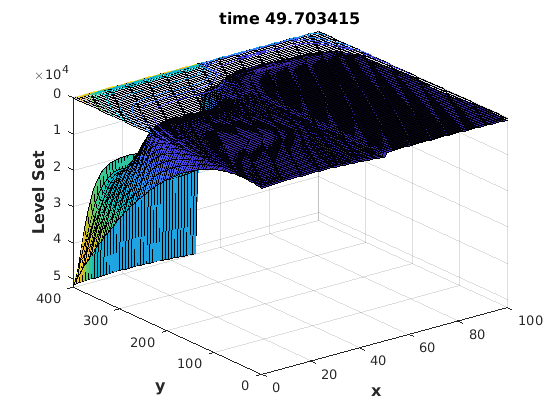}
\includegraphics[width=4.0cm,angle=-0]{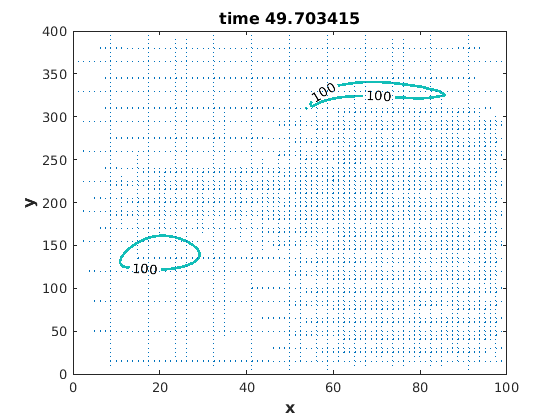} 
\end{center}
\caption{\label{burger_E_de} Upper figures: Decoupled velocity-field of the Burgers' equation with weak E-field, which is computed in a maximum of the amplitudes (contour visualization, left figure), right figure: Velocity-field of the Burgers' equation with weak E-field (vectorial-visualization, right figure). Lower Figures: Transport of 2 bubbles in a velocity field with decoupled velocity field computed by the Burgers' equation with an E-field, the level-set function is given in the lower left figure and the contour functions is given in the lower right figure at time $t=50$.}
\end{figure}

\end{itemize}

\subsection{Test example 3: Coupled EHD-model with Burgers' equation}

We apply the following coupled model, which is applied with $p = 0$ and we assume, that the velocity of the Burgers' equation are in the scales of the
Level-set equation, means we also have to apply Burgers' equation simultaneously to the Level-set equation, means we have $\Delta t_{LS} \approx \Delta_{Burgers}$.

\begin{figure}[ht]
\begin{center} 
\includegraphics[width=8.0cm,angle=-0]{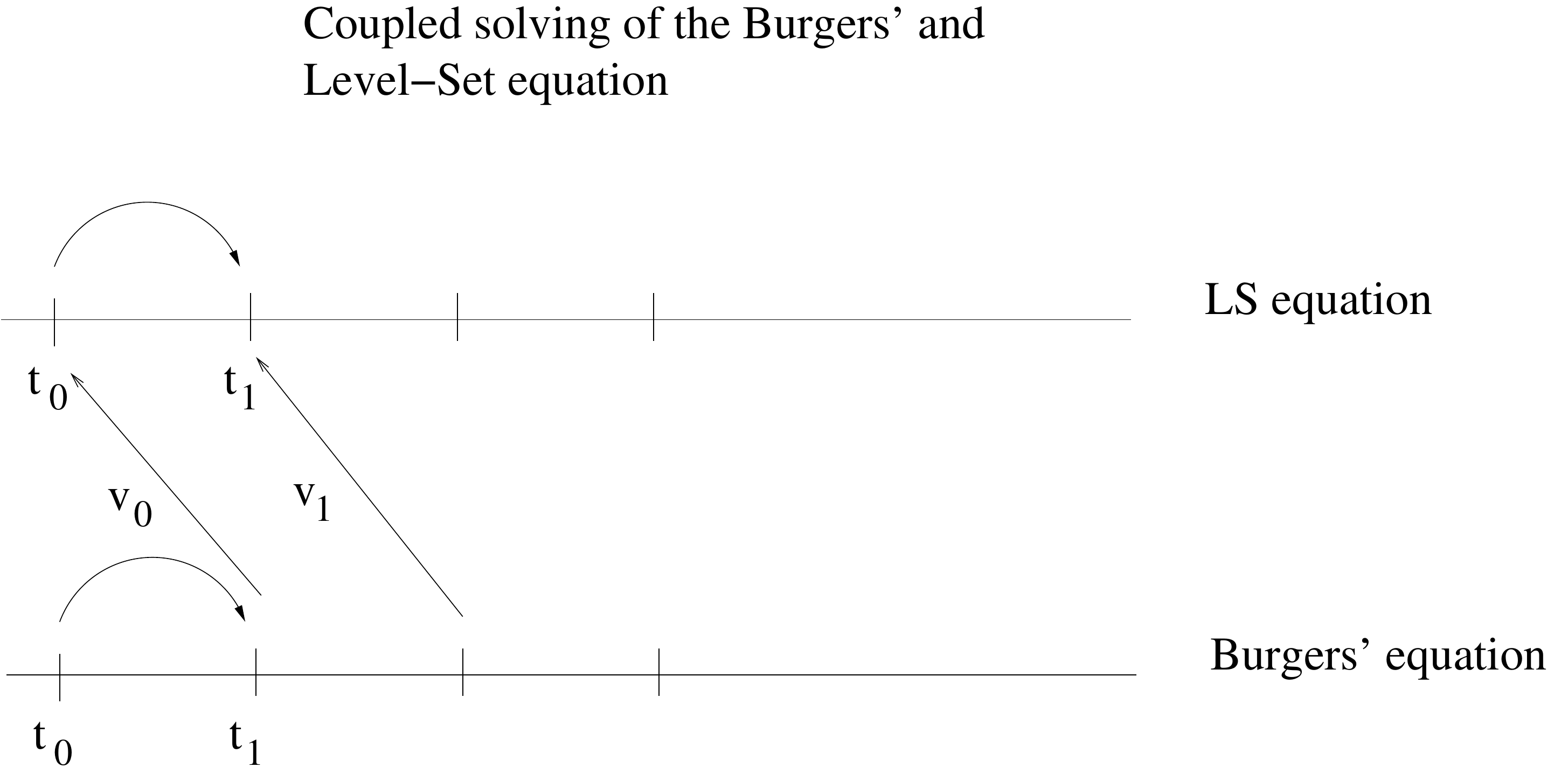}
\end{center}
\caption{\label{simult} Coupled velocity-field between the Burgers' and the LS equation (update in each simple time-step).}
\end{figure}

Here, we have a fast process that influences the bubbles, while the velocity field influences the speed of the bubbles in the field, we have to deal with the 
repeatedly update, see Figure \ref{simult}.

We have the Burgers' equation, which is given as:
\begin{eqnarray}
&& \rho \frac{\partial {\bf u}}{\partial t} + \rho ( {\bf u} \nabla ) {\bf u} = \mu \nabla^2 {\bf u} + {\bf f}_{ef}  , {\bf x} \in [0, 100] \times [0, 400], t \in [0, T_{end}] , \\
&& {\bf u}({\bf x}, 0) = \left\{
\begin{array}{cc}
1.5 , &  12.5 \le x \le 56.25 \; \mbox{and} \;   150.0 \le x \le 225.0 , \\
1.0 , & \mbox{else}
\end{array}
\right.
\end{eqnarray}
where ${\bf u}$ is the 2D-velocity field and $p$ is the pressure, $\mu$ is the dynamic viscosity, $\rho$ is the fluid density, ${\bf f}_{ef}$ is the electrostatic force.
We apply $\rho = 1.0$ and $\mu = 0.01$ and $T_{end} = 40$.

The forces are given as:
\begin{eqnarray}
&& {\bf f}_{ef} = \nabla \tau_e = \rho_e {\bf E} ,
\end{eqnarray}
where we have ${\bf E} ={\bf E}_1 + {\bf E}_2$ , \\
$ {\bf E}_1(x,y) = \left\{ \begin{array}{cc}
   (|\sin(2 \pi t)|, |\sin(2 \pi t)|)^t, &  \mbox{for} \; (x,y) \in (0.45, 0.55) \times (0.45, 0.55) \\
  0 & \mbox{else}
\end{array} \right. $ and \\

$ {\bf E}_2(x,y) = \left\{ \begin{array}{cc}
  (|(\sin(2 \pi t)|, |\sin(2 \pi t)| )^t, &  \mbox{for} \; (x,y) \in (0.75, 0.85) \times (0.75, 0.85) \\
  0 & \mbox{else}
\end{array} \right. $ \\
we assume only positive E-fields.

Further, we have the CFL-condition of the Burgers' Equation and the 
level-set equation for the
time-interval $[t^n, t^{n+1}]$ is given as:

\begin{itemize}
\item CFL-Condition for the Burgers' equation:
\begin{eqnarray}
&& \Delta t \le \frac{1}{\frac{\max\{{\bf u}^n\}}{\Delta x} + \frac{2 D }{\Delta x}} ,
\end{eqnarray}
where $\Delta t$ is the time-step, $\Delta x = \Delta y$ is the spatial step and
$\max\{{\bf u}^n\}$ is the absolute value of the velocity-field in time-step $t^n$ and $D$ is the diffusion parameter.
\item CFL-Condition for the Burgers' equation:
\begin{eqnarray}
&& \Delta t \le \frac{1}{\frac{\max\{{\bf u}^n\}}{\Delta x} + \frac{2 D }{\Delta x}} ,
\end{eqnarray}
where $\Delta t$ is the time-step, $\Delta x = \Delta y$ is the spatial step and
$\max\{{\bf u}^n\}$ is the absolute value of the velocity-field in time-step $t^n$ and $D$ is the diffusion parameter.

\end{itemize}

The velocity field is applied in the phase field model, which is given as:
\begin{eqnarray}
\label{gas_phase_1}
&& \frac{\partial \tilde{c}_i}{\partial t} + \nabla({\bf u} \; \tilde{c}_i) = 0 ,  \;  (x,y, t) \in \Omega \times [0, T] , \; i = 1,  \ldots , M , \\
\label{gas_phase_2}
&& \tilde{c}_i({\bf x}, 0) = \tilde{c}_{i,0}({\bf x}) , \\
\label{gas_phase_3}
&& i = 1,  \ldots , M ,
\end{eqnarray}
where ${\bf u}$ is the velocity of the two-phase flow equation, and $\tilde{c}_i$ are the indicator functions of the gas-phases
and $\tilde{c}_{i,0}({\bf x})$ are the initial conditions. The indicator functions $\tilde{c}_i({\bf x}, t)$ is given as:
\begin{eqnarray}
\left\{ \begin{array}{cc}
\tilde{c}_i({\bf x}, t) = 1 , & \mbox{gas-phase} , \\ 
\tilde{c}_i({\bf x}, t) = 0 , &  \mbox{liquid-phase} , \\ 
0 < \tilde{c}_i({\bf x}, t) < 1 , &  \mbox{interface} ,
\end{array} \right.
\end{eqnarray}
with $i = 1, \ldots, M$, while $M$ is the number of bubble-sources.

Further, we have the CFL-condition of the Burgers' Equation and the 
level-set equation for the
time-interval $[t^n, t^{n+1}]$ is given as:

\begin{itemize}
\item CFL-Condition for the Burgers' equation:
\begin{eqnarray}
&& \Delta t \le \frac{1}{\frac{\max\{{\bf u}^n\}}{\Delta x} + \frac{2 D }{\Delta x}} ,
\end{eqnarray}
where $\Delta t$ is the time-step, $\Delta x = \Delta y$ is the spatial step and
$\max\{{\bf u}^n\}$ is the absolute value of the velocity-field in time-step $t^n$ and $D$ is the diffusion parameter.

\item CFL-Condition for the Level-set equation:
\begin{eqnarray}
&& \Delta t \le \frac{\Delta x}{\max\{{\bf u}^n\}} ,
\end{eqnarray}
where $\Delta t$ is the time-step, $\Delta x = \Delta y$ is the spatial step and
$\max\{{\bf u}^n\}$ is the absolute value of the velocity-field in time-step $t^n$.

\end{itemize}

\begin{itemize}

\item EHD-model without E-field, we have $\rho_e = 0.0$: \\

The velocity field is computed by the Burgers' equation, see the Figures \ref{burger_no_E} (here without the E-field). Further, we apply the computed velocity field based on the Burgers' equation (without the E-field) into the phase field model,
which is computed with modified level-set method, see the Figures \ref{burger_no_E}.
\begin{figure}[ht]
\begin{center} 
\includegraphics[width=4.0cm,angle=-0]{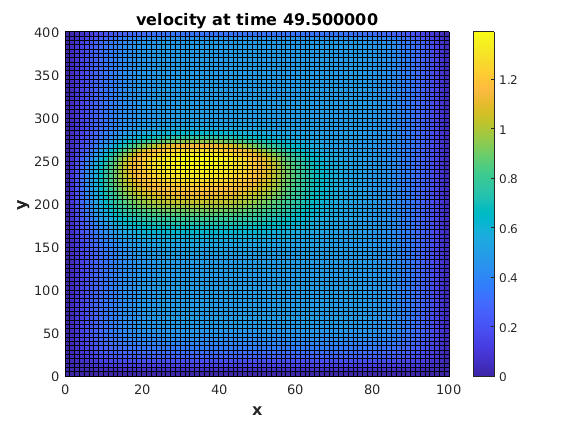}
\includegraphics[width=4.0cm,angle=-0]{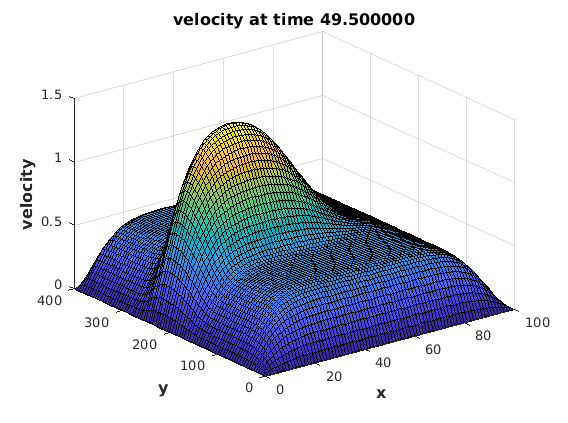} \\
\includegraphics[width=4.0cm,angle=-0]{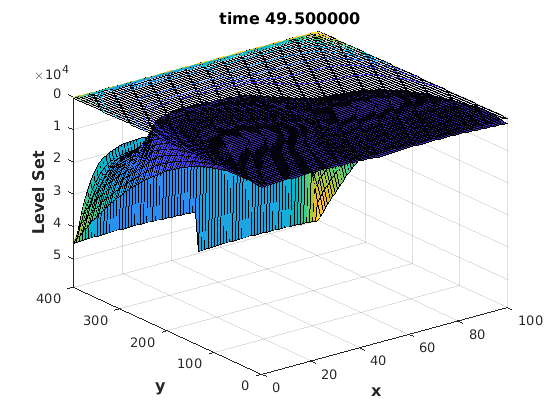}
\includegraphics[width=4.0cm,angle=-0]{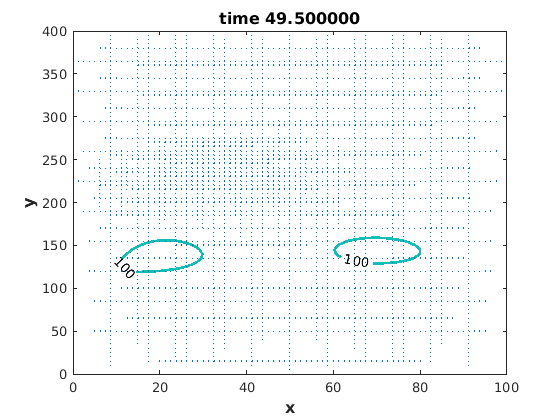} 
\end{center}
\caption{\label{burger_no_E} Upper figures: Coupled velocity-field of the Burgers' equation without E-field in contour visualization (upper left figure) and velocity-field of the Burgers' equation without E-field in vectorial-visualization (upper right figure). Lower Figures: Transport of 2 bubbles in a velocity field with coupled velocity field computed by the Burgers' equation without an E-field, the level-set function is given in the left lower figure and the contour functions is given in the lower right figure at time $t=50$.}
\end{figure}

\item EHD-model with E-field:, we have $\rho_e = 0.01$ \\

The velocity field is computed by the Burgers' equation, see the Figures \ref{burger_E} (here with the E-field). Further, we apply the computed velocity field based on the Burgers' equation (with the E-field) into the phase field model,
which is computed with modified level-set method, see the Figures \ref{burger_E}.
\begin{figure}[ht]
\begin{center} 
\includegraphics[width=4.0cm,angle=-0]{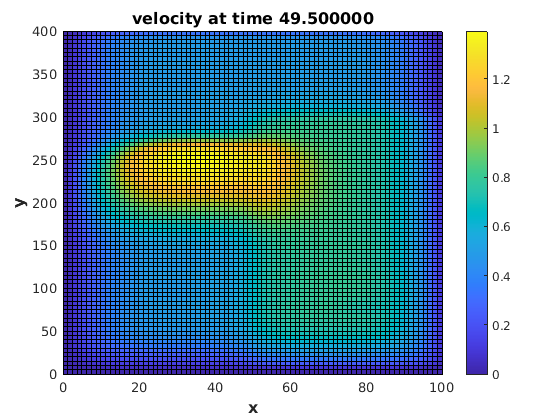}
\includegraphics[width=4.0cm,angle=-0]{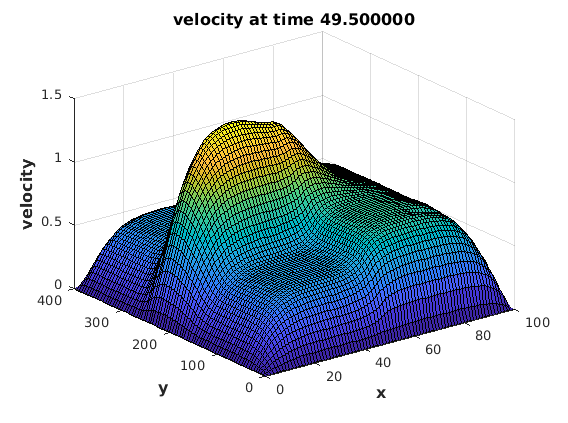} \\
\includegraphics[width=4.0cm,angle=-0]{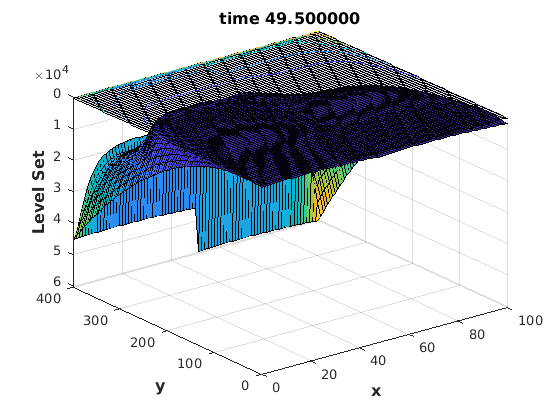}
\includegraphics[width=4.0cm,angle=-0]{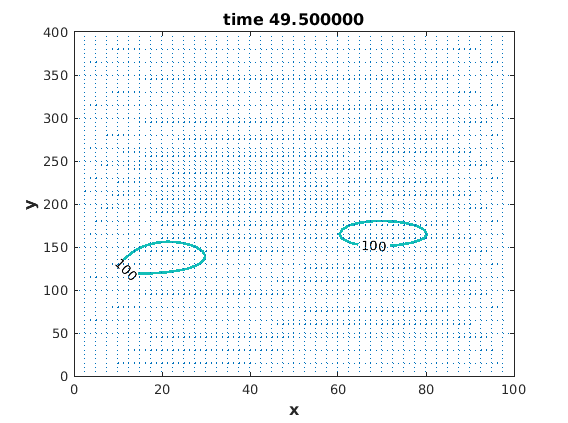} 
\end{center}
\caption{\label{burger_E} Upper figures: Coupled velocity-field of the Burgers' equation with weak E-field, which is computed by each time-step, in contour visualization (upper left figure) and velocity-field of the Burgers' equation with weak E-field in vectorial-visualization (upper right figure). Lower Figures: Transport of 2 bubbles in a velocity field with coupled velocity-field computed by the Burgers' equation with an E-field, the level-set function is given in the lower left figure and the contour functions is given in the lower right figure at time $t=50$.}
\end{figure}

\end{itemize}

\section{Conclusion}
\label{concl}

In this paper, we discussed coupled models, which are based on electro-hydrodynamics equations and transport models, which are based on phase-field equations.
The models are coupled via splitting approaches and we could present decoupled and coupled versions of the electrohydrodynamical model with the
phase field model.
We obtained numerical results, while we have to be careful with the different scales of the bubble-oscillations in the E-field and the
flow-scale if the hydrodynamical equation. We solved such problems with time- and space-step controls. 
First numerical results are presented with the coupled EHD and phase-field models, here we could see the influence of the different and changing velocities, due to the E-Field on the
bubbles.

\bibliographystyle{plain}

\end{document}